\documentclass[conference]{IEEEtran}
\ifCLASSINFOpdf
\else
\fi
\usepackage{color}
\usepackage{enumerate}
\usepackage[cmex10]{amsmath}
\usepackage{siunitx}
\usepackage{import}
\usepackage{enumitem}
\usepackage{indentfirst}
\usepackage{standalone}
\usepackage[section]{placeins}
\usepackage{pdflscape}
\usepackage{tabularx}
\usepackage{algorithm}
\usepackage{amssymb}
\usepackage{microtype}
\usepackage{soul}
\usepackage{mathtools}
\usepackage{multirow}
\usepackage{verbatim}
\usepackage{graphicx}
\usepackage{pgfplots}

\usepackage{subfloat}
\usepackage{verbatim}
\usepackage[utf8]{inputenc}
\usepackage[T1]{fontenc}
\usepackage{amsmath}
\usepackage[space]{grffile}
\usepackage{booktabs}
\usepackage{xfrac}
\usepackage{tabularx}
\usepackage{cite}
\usepackage{adjustbox}
\usepackage{verbatim}
\usepackage{url}
\usetikzlibrary{shapes}
\usepackage{multicol, blindtext}
\usepackage[]{color}
\usepackage{framed}
\usepackage{alltt}
\usepackage[T1]{fontenc}
\usepackage{array}
\usepackage{colortbl}
\usepackage{booktabs}
\usepackage{multirow}
\usepackage{eurosym}
\usepackage{listings}
\usepackage{ifthen}
\usepackage{pgfplots}
\usepackage{pgfplotstable}
\usepackage{pgfcalendar}
\usepackage{pgfgantt}
\usepackage{tikz}
\usetikzlibrary{shapes.geometric, arrows}
\usepackage{longtable}
\usepackage{amsfonts}
\usepackage{subfigure}
\usepackage{float}
\usepackage{mathrsfs}
\usepackage{algpseudocode}
\usepackage{textcomp}
\usepackage{xspace}
\makeatletter
\newcommand\fs@betterruled{
  \def\@fs@pre{\vspace*{5pt}\hrule height.8pt depth0pt \kern2pt}\def\@fs@post{\kern2pt\hrule\relax}\def\@fs@mid{\kern2pt\hrule\kern2pt}\let\@fs@iftopcapt\iftrue}
\floatstyle{betterruled}
\restylefloat{algorithm}
\makeatother
\pgfplotsset{compat=1.16}
\usepackage{fmtcount}

\usepackage[
  top=0.72in,
  bottom=1in,
  left=0.64in,
  right=0.64in
]{geometry}
\begin{document}

\title{A Low-Latency Semantic State Estimator using Latent Predictive Learning for Dynamic Network Monitoring and Orchestration
\vspace{-0.3cm}
}

	\author{\IEEEauthorblockN{
			\parbox{\linewidth}{\centering
                  Hari Madhukumar\IEEEauthorrefmark{1},
                  Haiyuan Li\IEEEauthorrefmark{1},
                  Xiaolan Liu\IEEEauthorrefmark{1},
                  Andy Corston-Petrie\IEEEauthorrefmark{6},
                  Dimitra Simeonidou\IEEEauthorrefmark{1}
			}
            \IEEEauthorblockA{\textit{Smart Internet Lab, University of Bristol\IEEEauthorrefmark{1}, BS8 1UB, U.K.},
            \textit{BT Group\IEEEauthorrefmark{6}, IP5 3RE, U.K.}\\
            E-mail: \{h.madhukumar, ocean.h.li, xiaolan.liu, dimitra.simeonidou\}@bristol.ac.uk,} andy.petrie@bt.com
            \vspace{-0.1cm}
		}
	}

\maketitle

\begin{abstract}
Closed-loop network monitoring and orchestration increasingly require semantic interpretations of live telemetry beyond raw counter collection.
However, dynamic cloud-edge environments change both the active node set and the monitoring query at runtime, while control loops demand bounded millisecond-scale responses. We introduce a latent predictive state estimator (LPSE) for dynamic network monitoring and orchestration, built on latent predictive learning over streaming telemetry. The framework converts variable-cardinality node telemetry into topology-adaptive temporal representations, fuses them with monitoring questions, and returns bounded answers from a semantic codebook instead of autoregressive text generation. This design enables fixed-cost, single-pass inference while preserving semantic interpretability. By operating on permutation-invariant, slot-routed node representations keyed by stable identity,
the model maintains a fixed input space and generalizes to node addition, removal, and reordering without retraining.
Experimental results on a multi-node Kubernetes cluster show semantic prediction accuracy of 87.13\% at approximately 72$\times$ lower mean inference latency compared to the lowest latency LLM tested.
\end{abstract}


\section{Introduction}

Future network infrastructures will rely on closed-loop monitoring and orchestration to maintain service quality under fluctuating traffic demands, service requirements, and resource conditions. In cloud-edge and virtualized environments, orchestration systems continuously collect network telemetry from distributed infrastructure nodes to support service placement, scaling, fault recovery, and admission-control decisions.
This demand intensifies as network operation evolves toward multi-agent AI orchestration~\cite{Li_Madhukumar_Yan_Wu_Simeonidou_2026, Jiang_AgenticIBN_2026}, where concurrent reasoning agents issue semantic state queries to monitoring systems within bounded response budgets. In single-query settings, hand-tuned threshold rules and dashboards remain competitive. The case for a learned bounded-output semantic estimator emerges when an LLM-based orchestrator would otherwise become a per-query bottleneck.

Semantic monitoring \cite{MartinezCasanueva_SemanticTelemetry_2023} in dynamic network infrastructures remains challenging as both telemetry structure and monitoring intent vary at runtime. Active infrastructure nodes may expand, contract, or change order as services are instantiated, migrated, terminated, or recovered, causing fixed-position telemetry vectors to become semantically inconsistent after topology or service-state changes. Moreover, monitoring is not restricted to a single predefined prediction target. The same snapshot may also be queried under several different operational intents, so a practical model must jointly encode live multi-node telemetry and query-specific intent while producing stable, low-latency outputs suitable for closed-loop automation.

Prior approaches to deriving operational states from network telemetry fall into three groups.
(i) \textit{Rule-based monitoring systems}, such as threshold alarms and manually configured dashboards, provide interpretable signals for predefined conditions but require continuous rule maintenance as services, metrics, and node states change. Prometheus is the state-of-the-art for cloud-native cluster monitoring.
(ii) \textit{Lightweight AI-based predictors}, including statistical models and compact neural networks for anomaly detection~\cite{Jin_GNN4TS_2024} and resource-state classification~\cite{Saxena_CloudWorkload_2023}, offer low inference latency but are conventionally deployed on fixed telemetry layouts and fixed prediction targets. (iii) \textit{LLM-based telemetry reasoning}~\cite{Zhang_LLM4AIOps_2025} can process heterogeneous metric contexts and monitoring queries without task-specific output heads, but depends on autoregressive decoding.
All three groups conventionally assume fixed input structures that break under topology change and fixed output objectives that cannot serve multiple query types. LLM-based methods additionally incur autoregressive latency incompatible with control-loop budgets.

Recent advances in self-supervised learning (SSL) show that latent predictive objectives produce robust and transferable representations for complex environments \cite{lecun2022path,grill2020bootstrap,Assran_IJEPA_2023}. In network-oriented settings, low-bandwidth operation, temporal prediction, and task-relevant representations are emphasized over raw-signal transport~\cite{Girgis_Valcarce_Bennis_2025,Monemi_Chinipardaz_Rasti_Bennis_Latva-Aho_2025}, with extensions to sequence modeling, graphs, and control~\cite{Skenderi_Li_Tang_Cristani_2025}.
However, most existing evidence remains simulation-based or benchmark-driven, whereas practical deployment exposes non-stationary telemetry distributions, runtime topology changes, online supervision without manual labels, and strict end-to-end latency bounds.
In response, we design a hybrid semantic state estimator for continuous network monitoring and orchestration in dynamic infrastructure.

\subsubsection{Question-conditioned semantic monitoring} We formulate closed-loop semantic monitoring as a question-conditioned state estimation problem over a variable-cardinality node set. The formulation models a dynamic telemetry input and a query-conditioned output drawn from a bounded semantic inventory, subject to dynamic node ordering and a millisecond-scale latency constraint. It is a transferable contribution and confers the same topology robustness on LSTM and Transformer estimators trained on the same tokens.

\subsubsection{Topology-adaptive telemetry representation} We propose a node-token telemetry representation that maps a variable set of live Kubernetes nodes to a fixed-dimensional cluster state using identity-keyed slot routing invariant to node ordering. Stable node identities are kept for question grounding and answer rendering, so the representation remains valid when nodes are added, removed, or reordered.

\subsubsection{Question-conditioned semantic estimation} Building on this representation, we develop a latent predictive semantic estimator combining temporal telemetry representations with structured monitoring questions and predicts answers from a fixed semantic codebook, enabling one model to handle multiple query types without autoregressive text generation.

\subsubsection{Live deployment and low-latency validation} We deploy the model on a live multi-node Kubernetes cluster and construct question-answer supervision automatically from network telemetry. The proposed model strikes a balance between performance and robustness at approximately 72$\times$ lower mean latency than a deployable 4B endpoint, making it suitable for embedding inside multi-agent orchestrators over LLMs where per-query LLM decoding would otherwise dominate end-to-end latency.

\section{System Model \& Problem Formulation}
\label{sec:problem_formulation}

We consider a live monitoring setting in which a service observes a changing set of compute nodes. Let $\mathcal{N}_t$ denote the set of active nodes observed at time $t$, indexed by $n$. Since nodes may join, leave, or recover over time, $|\mathcal{N}_t|$ is not fixed. Each node reports telemetry from multiple resource and system-level metric families. Let $\mathcal{M}$ denote the set of monitored metric families, indexed by $m$. For node $n \in \mathcal{N}_t$ and metric family $m \in \mathcal{M}$, the samples collected at time $t$ are summarized as a fixed-dimensional descriptor
$\mathbf{r}_{n,m}(t)\in\mathbb{R}^{D_r}$.

The telemetry state of node $n$ at time $t$ is represented by the node token
\begin{equation}
    \mathbf{v}_{n}(t)=
    [\mathbf{r}_{n,m_1}(t);\mathbf{r}_{n,m_2}(t);\ldots;\mathbf{r}_{n,m_{|\mathcal{M}|}}(t)]
    \in \mathbb{R}^{F_{\text{node}}},
\end{equation}
where $F_{\text{node}}=D_r|\mathcal{M}|$. Each node is also associated with a stable identity $u_n$. The telemetry snapshot at time $t$ is then written as the variable-cardinality set
\begin{equation}
    \mathcal{V}_t=\{(\mathbf{v}_{n}(t),u_n)\mid n\in\mathcal{N}_t\}.
    \label{eq:telemetry_snapshot}
\end{equation}
This set representation does not assign semantic meaning to node order and naturally supports changes in the number of observed nodes.

The monitoring service is queried by an operator or orchestration agent. Let $q_t$ denote the query at time $t$. A query specifies an operational intent, such as pressure detection, node comparison, bottleneck identification, condition checking, or short-horizon trend estimation. It may also include a metric scope, a node reference, or a time horizon. The answer space is a finite semantic inventory $\mathcal{A}=\{a_1,\ldots,a_\kappa\}$, where each $a_k$ denotes an operational answer type. During training, each pair of telemetry history and query is assigned a target answer $a_t^\star\in\mathcal{A}$.

The estimator takes a temporal history of telemetry snapshots $\mathcal{H}_t = [\mathcal{V}_{t-T+1}, \ldots, \mathcal{V}_t]$, where $T$ is the context-window length. Given $\mathcal{H}_t$ and $q_t$, the estimator predicts a categorical distribution $p_\theta(\cdot \mid \mathcal{H}_t, q_t)$ over $\mathcal{A}$ parameterised by $\theta$, and returns
\begin{equation}
    \hat{a}_t =
    \arg\max_{a_k\in\mathcal{A}}
    p_{\theta}(a_k\mid \mathcal{H}_t,q_t),
    \label{eq:semantic_estimator}
\end{equation}
where $\theta$ denotes the model parameters. Let $\ell_{\theta}(\mathcal{H}_t,q_t)$ be the inference latency, let $\tau_{\text{ctrl}}$ be the response-time budget of the closed loop, and let $\pi(\cdot)$ denote any permutation of node order within each snapshot's tensor representation. Let $q'\sim q_t$ denote that $q'$ is a semantically equivalent rephrasing of $q_t$. The semantic monitoring problem is formulated as
\begin{subequations}
\label{eq:semantic_problem}
\begin{align}
    \min_{\theta} \quad &
    \mathbb{E}_{t}\!\left[-\log p_{\theta}(a_t^\star\mid \mathcal{H}_t,q_t)\right]
    \label{eq:semantic_obj}\\
    \text{s.t.}\quad
    & p_{\theta}(\cdot\mid \pi(\mathcal{H}_t),q_t)
      =
      p_{\theta}(\cdot\mid \mathcal{H}_t,q_t), \quad \forall \pi,
      \label{eq:perm_inv}\\
    & \hat{a}_t \in \mathcal{A},
      \label{eq:bounded_answer}\\
    & \ell_{\theta}(\mathcal{H}_t,q_t) \le \tau_{\text{ctrl}},
      \label{eq:latency_constraint}\\
    & p_{\theta}(\cdot\mid \mathcal{H}_t,q_t)
      \approx
      p_{\theta}(\cdot\mid \mathcal{H}_t,q'), \quad \forall\, q'\sim q_t.
      \label{eq:paraphrase_inv}
\end{align}
\end{subequations}
The objective in Eq.~\eqref{eq:semantic_obj} fits the estimator to the target semantic answer (the point target $a_t^\star$ is relaxed to a soft-label distribution at training time; see \S{}\ref{sec:training_objective}). Constraints~\eqref{eq:perm_inv}--\eqref{eq:bounded_answer} enforce invariance to node ordering, which with the set-based snapshot also admits node addition and removal, and confine the prediction to the semantic inventory. Constraint~\eqref{eq:latency_constraint} imposes the closed-loop response-time (50ms). Constraint~\eqref{eq:paraphrase_inv} expresses the corresponding invariance on the query side. Semantically equivalent phrasings should yield approximately the same answer distribution. Unlike Eq.~\eqref{eq:perm_inv}, which the slot-routed representation of \S{}\ref{sec:framework} satisfies by construction, this is stated as an approximation that the formulation is designed toward rather than a guaranteed property.

\section{LPSE Framework}
\label{sec:framework}
\begin{figure*}[h!]
    \centering
    \includegraphics[width=1\linewidth]{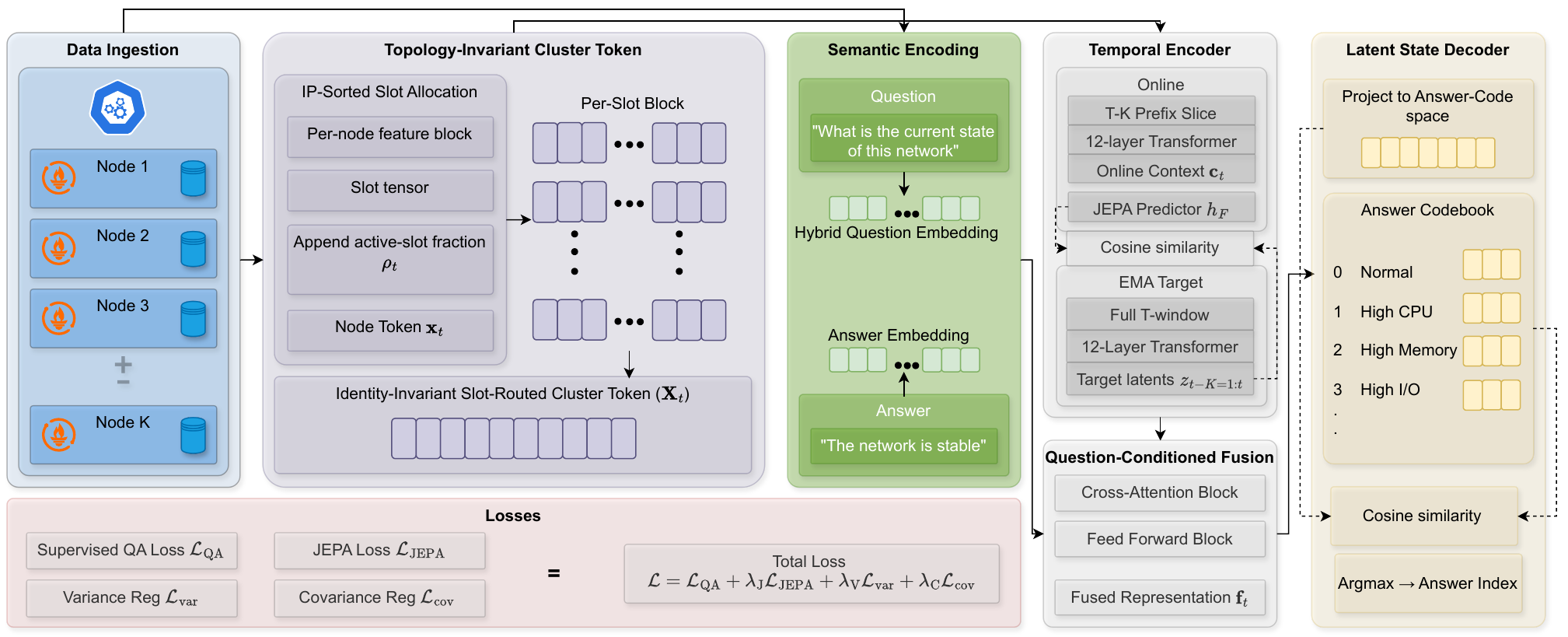}
    \caption{Pipeline for dynamic Kubernetes monitoring that includes multi-node metric ingestion, temporal latent encoding, question-conditioned latent fusion, codebook-based semantic answering, and confidence-aware output selection.}
    \label{fig:jepa-example}
    \vspace{-0.3cm}
\end{figure*}

At each sampling step, node-level metrics are collected across all discovered cluster nodes. The pipeline from data ingestion to answer selection is shown in Fig.~\ref{fig:jepa-example}. The framework computes six summary statistics per tracked metric family (count, minimum, maximum, mean, sum, and standard deviation), instantiating the per-node descriptor $\mathbf{v}_n(t) \in \mathbb{R}^{F_{\text{node}}}$ and the variable-size node-token set $\mathcal{V}_t$ \eqref{eq:telemetry_snapshot}, with $u_n$ given by the node's IP address. Instead of flattening along an arbitrary node axis, LPSE routes each node into a fixed slot inside an $S$-slot cluster tensor. Slots are assigned in IP-sorted rank order of $u_n$ and higher-ranked nodes get shifted by one slot when nodes arrive or leave. Each occupied slot stores its $F_{\text{node}}$ feature block, a four-dimensional normalised IP identity vector, and a one-dimensional presence flag. Empty slots are zero-padded. The cluster token also carries a final scalar $\rho_t = \min(|\mathcal{N}_t|, S)/S$ holding the active-slot fraction:
\begin{equation}
\mathbf{x}_t = [\,\mathrm{slot}_1;\,\mathrm{slot}_2;\,\ldots;\,\mathrm{slot}_S;\,\rho_t\,] \in \mathbb{R}^{F_{\text{cluster}}},
\end{equation}

where $F_{\text{cluster}} = S\,(F_{\text{node}} + 5) + 1$. Because each slot also carries the IP identity of whichever node currently occupies it, reordering the input set leaves $\mathbf{x}_t$ unchanged: identical multisets of (features, IP) pairs always yield the same cluster token. This satisfies the permutation-invariance requirement of Eq.~\eqref{eq:perm_inv} via stable identity rather than the symmetric pooling used by Deep Sets~\cite{Zaheer_DeepSets_2017} and Set Transformer~\cite{Lee_SetTransformer_2019}, and it keeps the input dimensionality constant as $|\mathcal{N}_t|$ varies up to the slot capacity $S$. Identity remains with the slot contents rather than the slot index, and the encoder is trained to read it from the per-slot IP vector and presence flag, so a single checkpoint remains valid across topology changes.

A temporal context window $\mathbf{X}_t = [\mathbf{x}_{t-T+1},\ldots,\mathbf{x}_t] \in \mathbb{R}^{T \times F_{\text{cluster}}}$ is maintained, and each feature is normalised online using running mean/variance statistics. The normaliser is applied to the slot-routed features only and per-slot identity entries pass through unscaled, since they contain routing information.

\subsection{Metric Encoder and Question-Conditioned Fusion}

The metric stream is processed by a Transformer-based temporal encoder with sinusoidal positional encoding. This module outputs (i) a full latent sequence $\mathbf{Z}_t = \text{Enc}_{\theta}(\mathbf{X}_t), \quad$ and (ii) a context summary vector $\mathbf{c}_t = \mathbf{Z}_t^{\text{last}}.$

In parallel, each natural-language query is converted into a structured intent representation (intent, metric, scope, horizon, node reference) and embedded into a fixed-dimensional question vector. A cross-attention fusion module then applies question-as-query attention over temporal metric latents $\mathbf{f}_t = \text{Fuse}_{\phi}(\mathbf{Z}_t, \mathbf{q}_t),$ yielding a question-conditioned state representation for answer selection.

\subsection{Semantic Codebook Output Space}

Instead of autoregressive decoding, LPSE uses a learnable answer codebook of $\kappa$ code vectors, each tied to a fixed natural-language template. The fused representation is projected into the answer-code space and compared to all code vectors using temperature-scaled cosine similarity:
\begin{equation}
s_k = \frac{\langle \hat{\mathbf{f}}_t, \hat{\mathbf{a}}_k \rangle}{\tau}, \quad k \in \{1,\ldots,\kappa\},
\end{equation}
where $\hat{\mathbf{f}}_t$ and $\hat{\mathbf{a}}_k$ are the L2-normalised fused state and code vectors and $\tau = 0.07$. The highest-scoring index determines the output template, which is subsequently filled with current node-level values (for example, node IDs or measured percentages) to produce a human-readable response.

\subsection{Training Objective and Continual Adaptation}
\label{sec:training_objective}

LPSE is trained online from live telemetry. Supervision is generated automatically through a QA synthesizer that transforms semantic metric features into question-template and answer-index pairs, eliminating manual annotation. This makes the model self-updating with respect to the live cluster rather than a static dataset.

The total loss combines a supervised QA objective, a JEPA-style self-supervised objective, and a pair of VICReg-style anti-collapse regularisers on the QA latent:
\begin{equation}
\mathcal{L} = \mathcal{L}_{\text{QA}} + \lambda_{\text{J}}\mathcal{L}_{\text{JEPA}} + \lambda_{\text{V}}\mathcal{L}_{\text{var}} + \lambda_{\text{C}}\mathcal{L}_{\text{cov}}.
\end{equation}

$\mathcal{L}_{\text{QA}}$ is a soft-label cross-entropy: the true template (index $k^\star$) keeps mass $1-\alpha = 0.70$, with the remaining $\alpha$ distributed over same-group templates $j$ in proportion to $e^{-|j-k^\star|/\tau_s}$, where $\tau_s$ is a spread temperature, plus optional uniform smoothing. $\mathcal{L}_{\text{JEPA}}$ is a latent-prediction objective in the I-JEPA family with BYOL-style exponential-moving-average (EMA) targets. It aligns the online encoder with an EMA target encoder by predicting EMA target latents at a held-out future horizon from a strictly truncated prefix. $\mathcal{L}_{\text{var}}$ enforces a per-dimension standard-deviation floor on the question-conditioned answer latent, and $\mathcal{L}_{\text{cov}}$ penalises the squared off-diagonal entries of its covariance matrix. We retain the JEPA-style label for the auxiliary loss because it adopts the latent-prediction objective; the system as a whole adopts JEPA, BYOL, and VICReg as influences. Exact training-time details are given in \S{}\ref{sec:training-procedure}.

To improve coverage of rare operational states (CPU-bound, memory pressure, I/O contention, asymmetric load), cluster stress profiles are randomised during training, and samples are retained in a replay buffer so each optimisation step sees a diverse mix of historical and recent states. At inference time, a single forward pass returns top-$k$ candidate templates and confidence scores from the maximum codebook similarity.

\section{Implementation and Training}
\label{sec:training-procedure}
The system runs as a single service on a live Kubernetes cluster; training and inference share a single checkpoint but run on separate request queues, so live serving is never blocked by an optimisation step.

\subsubsection{Cluster Embedding}

Each \textit{node\_exporter} scrape is parsed into structured (family, label, value) tuples. For each family the framework computes six summary statistics, i.e., count, minimum, maximum, mean, sum, and standard deviation, yielding a 300-dimensional feature block per node ($D_r=6$, $|\mathcal{M}|=50$). These blocks are placed into a 16-slot cluster tensor in IP-sorted order; each occupied slot also carries a four-dimensional normalised IP vector and a presence flag, empty slots are zero-padded, and a final scalar records the active-slot fraction, giving a 4881-dimensional cluster token.

\subsubsection{Metric Encoder}

The encoder consumes a sliding window of $T=16$ cluster tokens, each linearly projected to 1024 dimensions, augmented with sinusoidal positional encoding, and processed by a 12-layer Transformer with 16 attention heads, GELU activations, pre-normalisation, and a feed-forward width of 4096. The most recent timestep's latent is the context summary supplied to the QA path, while the full sequence is retained for the JEPA self-supervised path.

\subsubsection{Question Encoder and Fusion}

We use a hybrid encoder with two parallel paths. A frozen SentenceTransformer \cite{Reimers_SBERT_2019} (\texttt{all-MiniLM-L6-v2}) produces the
sentence embedding, while an explicit identity path passes the four IP
octets through a small MLP and looks up the scope label in an embedding
table. The concatenation is projected to a 256-dimensional question
vector. The fusion module is a four-layer cross-attention stack (16
heads, 1024 dimensions) where the question forms the query and the
encoder sequence forms keys and values.

\subsubsection{Answer Codebook}

The output stage is what makes inference deterministic and bounded. Rather than decoding free-form text, the fused state is projected into a fixed inventory of $\kappa = 94$ semantic templates, and the index of the closest template is emitted as the answer. The templates are organised into thirteen operational groups (CPU level, memory level, network state, disk and filesystem fill, load, processes, temperature, cluster-wide health, pairwise node comparison, extreme-node identification, short-horizon prediction, condition checks, and a small generic fallback). This grouping is what makes the soft-label cross-entropy described below well-defined, since probability mass can be shared between adjacent CPU-level bands but not between unrelated groups. The inventory is built offline from the intents an operator wishes to expose, with candidate phrasings enumerated by an LLM and deduplicated. Therefore, it is deployment-specific, and a new domain requires its own enumeration and a retrained codebook projector.

\subsubsection{Online Supervision and Training Objective}

A live cluster cannot be hand-labelled at the rate at which it generates data. Supervision is therefore produced online: raw counters are turned into interpretable quantities (e.g., CPU utilisation, memory used fraction, load-per-core ratio, filesystem fill, TCP and process states), and fixed rules emit the appropriate answer-template index. For example, mean CPU utilisation is bucketed into seven levels with thresholds at 1\%, 10\%, 30\%, 50\%, 70\%, and 90\% (idle through critical). The model's purpose is not to discover them but to evaluate them jointly across the 16-step temporal context, multiple query types, and variable topology at bounded latency. Up to ten pairs are produced per snapshot and written to a replay buffer and one in every ten snapshots is diverted to a held-out validation buffer.

$\mathcal{L}_{\text{JEPA}}$ ($\lambda_{\text{J}} = 0.5$, with $\lambda_{\text{V}} = 1.0$ and $\lambda_{\text{C}} = 0.005$ for the VICReg terms) provides the self-supervised signal that prevents the encoder from collapsing onto features that minimise the QA loss alone, following the online/EMA prediction-in-latent-space paradigm of I-JEPA~\cite{Assran_IJEPA_2023} but with a temporal rather than spatial conditioning variable. Two encoders are maintained: an online encoder receiving gradients, and an EMA target encoder updated as $\theta_{\text{EMA}} \leftarrow m\,\theta_{\text{EMA}} + (1 - m)\,\theta_{\text{online}}$ with $m = 0.996$. For a window of length $T$, the online encoder is run on the $(T - K)$-prefix to obtain a context latent $\mathbf{c}^{\text{short}}_t$, the EMA encoder on the full window to obtain target latents $\bar{\mathbf{z}}_{t-K+1:t}$, and a small predictor $h_F(\cdot)$ maps the prefix context to a predicted latent compared to each target step in normalised cosine distance:
\begin{equation}
\mathcal{L}_{\text{JEPA}} = \frac{1}{K}\sum_{i=1}^{K} \left(1 - \cos\!\big(h_F(\mathbf{c}^{\text{short}}_t),\,\bar{\mathbf{z}}_{t-K+i}\big)\right).
\end{equation}
Here $K$ is the latent-prediction horizon of the auxiliary objective. $\mathcal{L}_{\text{var}}$ and $\mathcal{L}_{\text{cov}}$ are the standard VICReg variance-floor and off-diagonal covariance penalties, applied to the question-conditioned answer latent rather than to the encoder output, so that the fusion module and codebook projector cannot satisfy the QA loss by collapsing the $d_a = 512$ answer dimensions onto a low-rank subspace. The optimiser is AdamW (learning rate $10^{-3}$, weight decay $10^{-5}$, gradient clip norm $1.0$). Each step samples a batch of $B = 32$ tuples and updates the metric encoder, fusion, codebook, learnable parts of the question encoder, and the JEPA predictor jointly.

\subsubsection{Online Adaptation}

The key augmentation is bounded node dropout, applied \emph{before} the cluster token is built and \emph{before} QA generation runs. Because the live IP list, per-node tensors, and parsed metric families are all filtered by the same keep-mask, the cluster token, semantic features, and QA labels agree on which nodes exist for that step, so node-referenced questions only mention IPs the model actually sees. Combined with the IP-sorted slot routing introduced above, this is what trains the encoder to read identity from the per-slot IP feature rather than from slot position.

\section{Experimental Results}
\label{sec:experiments}

\begin{figure}[t]
    \centering
    \setlength{\abovecaptionskip}{0cm}
    \subfigure[Joint loss]{
        \label{fig:loss_curve}
        \includegraphics[width=0.478\linewidth]{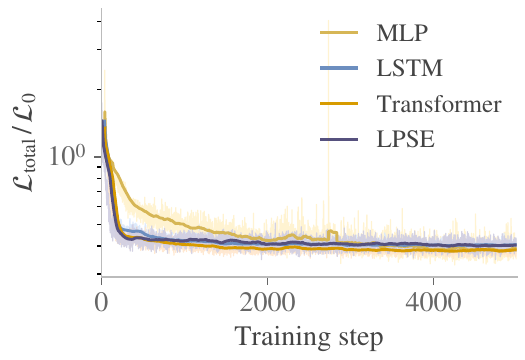}}
    \subfigure[Validation top-1 vs.\ step]{
        \label{fig:accuracy_curves}
        \includegraphics[width=0.478\linewidth]{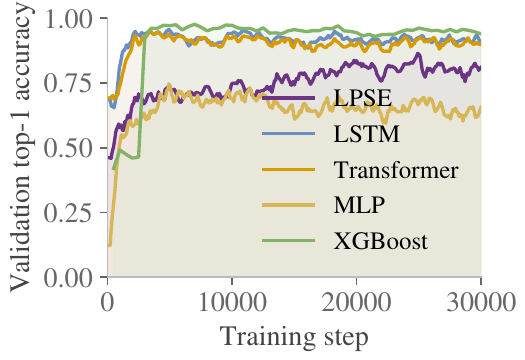}}

    \vspace{-0.2cm}

    \subfigure[End-of-run accuracy]{
        \label{fig:compute_matched_accuracy_a}
        \includegraphics[width=0.478\linewidth]{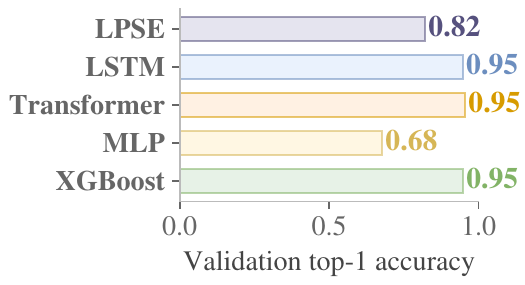}}
    \subfigure[Cumulative sample-views]{
        \label{fig:compute_matched_accuracy_b}
        \includegraphics[width=0.478\linewidth]{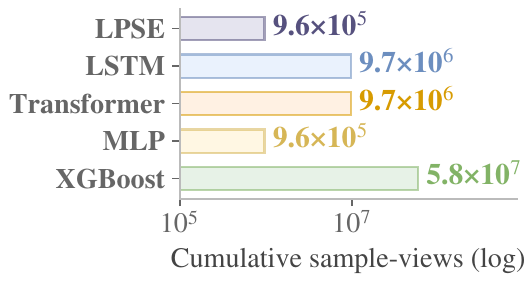}}

    \caption{Training dynamics (a, b) and end-of-run accuracy and compute (c, d).}
    \label{fig:training-results}
    \vspace{-0.3cm}
\end{figure}

\begin{figure}[t]
    \centering
    \setlength{\abovecaptionskip}{0cm}
    \includegraphics[width=0.8\linewidth]{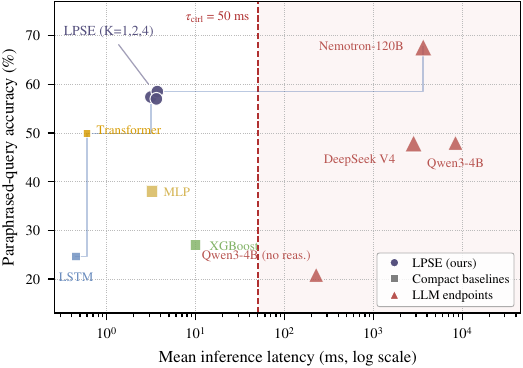}
    \caption{Accuracy-latency frontier on the 800-question benchmark.}
    \label{fig:pareto}
    \vspace{-0.3cm}
\end{figure}


\begin{table*}[t]
\centering
\caption{Cross-system comparison on the 800-question benchmark.}
\label{tab:jepa-vs-llm}
\scriptsize
\setlength{\tabcolsep}{6pt}
\renewcommand{\arraystretch}{1.0}
\resizebox{\textwidth}{!}{\begin{tabular}{@{}lcccccccc@{}}
\toprule
\textbf{System} &
\multicolumn{4}{c}{\textbf{Accuracy}} &
\multicolumn{4}{c}{\textbf{Cost per query}} \\
\cmidrule(lr){2-5}
\cmidrule(lr){6-9}
&
\textbf{Clean} &
\textbf{Drop one node} &
\textbf{Paraphrased} &
\textbf{Unique para.} &
\textbf{Mean lat. (ms)} &
\textbf{P95 lat. (ms)} &
\textbf{Memory (MB)} &
\textbf{Params} \\
\midrule
LPSE ($K{=}1$) &
\textbf{85.38\%} &
\textbf{84.6\%} &
\textbf{57.38\%} &
\textbf{53.8\%} &
\textbf{3.13} &
\textbf{3.19} &
\textbf{932} &
\textbf{240\,M} \\
LPSE ($K{=}2$) &
87.25\% &
84.8\% &
58.50\% &
57.7\% &
3.70 &
4.51 &
932 &
240\,M \\
LPSE ($K{=}4$) &
91.12\% &
89.6\% &
57.00\% &
53.8\% &
3.61 &
4.17 &
932 &
240\,M \\
LSTM &
93.75\% &
93.6\% &
24.62\% &
7.7\% &
0.45 &
0.44 &
24 &
6.1\,M \\
Transformer &
94.25\% &
92.9\% &
49.88\% &
38.5\% &
0.60 &
0.63 &
19 &
4.7\,M \\
MLP &
74.00\% &
38.00\% &
38.00\% &
39.06\% &
3.22 &
3.10 &
407 &
83.5\,M \\
XGBoost &
84.00\% &
60.00\% &
27.00\% &
29.69\% &
9.91 &
10.75 &
123 &
660 trees \\
Qwen3-4B (no reasoning) &
26.50\% &
29.2\% &
20.88\% &
16.7\% &
226.7 &
248.8 &
14{,}400 &
4\,B \\
Qwen3-4B &
55.12\% &
48.0\% &
48.0\% &
46.2\% &
8353 &
12606 &
14{,}400 &
4\,B \\
DeepSeek V4 Flash &
81.50\% &
73.6\% &
47.88\% &
69.2\% &
2824 &
4917 &
$\approx$186{,}000$^\dagger$ &
304\,B \\
Nemotron-120B &
80.75\% &
72.4\% &
67.62\% &
57.7\% &
3628 &
7765 &
$\approx$180{,}000$^\dagger$ &
120\,B \\
\bottomrule
\end{tabular}}
\vspace{0.2em}
\begin{minipage}{\textwidth}
\scriptsize
$^\dagger$ Total memory across the two RTX 6000 Pro Blackwell GPUs hosting the LLM, including weights and KV-cache.
\end{minipage}
\vspace{-0.3cm}
\end{table*}

LPSE reaches 87.13\% rolling validation accuracy under changing node configurations on the held-out validation buffer accumulated during training. An 800-question clean cross-system comparison (Table~\ref{tab:jepa-vs-llm}, 85.38\%) is reported above, primarily for paraphrased queries, latency and cost. The rolling accuracy is measured on the training-time validation buffer, whereas the cross-system benchmark uses fresh post-training snapshots, so the two are not directly comparable. In Table \ref{tab:jepa-vs-llm}, we compare results against LLM endpoints, a multilayer perceptron (MLP), an XGBoost classifier, and temporal predictors such as long short term memory (LSTM) and Transformer.

\subsection{Setup}
All evaluated systems are exercised against the same seven-node live Kubernetes cluster in a single benchmark run. Synthetic load is injected by periodically deploying \texttt{stress-ng} jobs to random subsets of workers under five profiles (CPU-bound, memory pressure, I/O contention, mixed, and asymmetric load), so the evaluator sees both naturally idle and actively stressed cluster states within the same run. All non-LLM systems are trained and evaluated on a single NVIDIA RTX 6000 Ada Generation GPU. The MLP and XGBoost baselines flatten the 16-frame context window with the question
embedding (78,352-dim input) through four 1,024-unit GELU layers
(83.5\,M parameters). In contrast, both temporal baselines (LSTM and Transformer) consume the \emph{full} 16-frame context window $\mathbf{x}_{t-15:t} \in \mathbb{R}^{B \times T \times F}$ \emph{unflattened}. All baselines share
our question encoder and codebook and train on the replay trace from LPSE's live run replayed chronologically so that every system observes the same stream in order. Implementations are released on Github\footnote{https://github.com/hari5g900/lpse}.

\subsection{Comparison: Accuracy, Robustness, and Cost}
\label{sec:cross-system}
Table~\ref{tab:jepa-vs-llm} consolidates the cross-system comparison, where every system answers the \emph{same} 800 questions on the \emph{same} sequence of fresh cluster snapshots. LPSE's 3.13\,ms mean latency leaves headroom for roughly fifteen sequential queries within a 50\,ms control-loop budget. Here, latency spans the full model path from tokenisation to template rendering but excludes telemetry collection, whose cost is a property of the deployment's scrape topology and is the same for all models. Fig.~\ref{fig:training-results} measures the training-time validation buffer, and the batch-fitted models transfer least well across that gap.

Under topology shift LPSE loses 0.78 points when a node is dropped, against the MLP's 36-point and XGBoost's 24-point drops. However, LSTM and Transformer, on the same slot-routed tokens, lose 0.15 and 1.35 points, so the robustness is a property of the identity-carrying representation and transfers across models. On clean queries at fixed topology the 4.7\,M-parameter Transformer is accordingly the stronger engineering choice at 94.25\% and 0.60\,ms. However, in multi-agent workflows, the advantage of the Transformer in clean-query accuracy must be weighed against the need for bounded and consistent semantic outputs. LPSE instead maps each query to a fixed semantic inventory, providing deterministic output selection.

A self-supervised ablation with the future-prediction term disabled but EMA consistency retained reaches only 64\% at step 30k (vs.\ 87\% for the full model) and plateaus at 74\% after $2.7\times$ more training. This suggests the importance of temporal asymmetry along with EMA consistency. $K$ is a deployment-time variable of that auxiliary objective, and its optimum depends on scrape cadence and workload volatility. We therefore use $K=1$ as the conservative default, while the results (Table~\ref{tab:jepa-vs-llm}) show that larger prediction horizons can improve clean and topology-shift accuracy, with diminishing or mixed benefits for paraphrased queries.

In a multi-agent orchestrator, wording variation is controlled by the caller agent with an LLM backend, so a robust model has to handle query-format drift. Since all baselines share LPSE's question encoder and codebook, the paraphrase column isolates the fusion and latent-predictive objective. Among the trained models, LPSE achieves the best paraphrase accuracy at 57.38\% ($K=1$), although it remains below the best LLM endpoint at 67.62\% (Fig. \ref{fig:pareto}). This gap highlights a remaining limitation in handling query-format variation; however, LPSE achieves this accuracy with substantially lower inference cost while maintaining bounded, deterministic outputs. This trade-off is particularly relevant to multi-agent orchestration, where multiple semantic queries may be issued within a single control-loop interval.

\section{Conclusion}

LPSE replaces autoregressive decoding with single-pass latent space prediction, reducing response time and confining outputs to a fixed operational inventory. On the evaluated multi-node Kubernetes cluster, the model achieved 87.13\% rolling validation accuracy under changing node configurations. At 3.13\,ms mean latency, it is 72$\times$ faster and 15$\times$ smaller than the lowest latency LLM tested (Qwen3-4B) while being more accurate in all benchmarks. LPSE supports multi-agent workflows, where low latency and robustness to query drift are the driving factors, and achieves the strongest overall performance among the evaluated models in this setting (Fig. \ref{fig:pareto}).

Several limitations remain. Paraphrase robustness drops to 57.38\% on reworded queries, the primary remaining gap to the strongest LLM tested, and on clean queries at fixed topology a far smaller Transformer is preferable. The slot routing fixes the input dimension at 16 nodes, so larger clusters require retraining at a wider slot capacity. The codebook is likewise deployment-specific. Future work will address paraphrase generalisation through fine-tuning or paraphrase-augmented training, incorporate richer cross-modal context such as logs and events, and couple the model with multi-agent orchestrators for proactive autonomous control.

\section*{Acknowledgment}
\small{This work was supported by the UK Engineering and Physical Sciences Research Council (EPSRC) grant EP/Y037243/1, EP/X04047X/2 for the TITAN Telecoms Hub and the Federated Telecoms Hubs, grant EP/Y036514/1 for the JOINER project, and the NVIDIA Academic Grant program.}

\bibliographystyle{IEEEtran}
\bibliography{references}

\end{document}